\documentclass[a4paper,traditabstract]{aa} 
\usepackage{graphicx} 
\usepackage{txfonts}
\usepackage{natbib}
\begin{document}

\title{Magnetic pattern at supergranulation scale: the Void Size Distribution }
\author{
	F. Berrilli\inst{1} 
	\and S. Scardigli \inst{1} 
	\and D. Del Moro \inst{1}
} 
\institute{Department of Physics, University of Rome Tor Vergata, I-00133, Rome, Italy  
                    \email{berrilli@roma2.infn.it}
          } 
\date{}
\abstract
{The large-scale magnetic pattern of the quiet sun is dominated by the magnetic network. 
This network, created by photospheric magnetic fields swept into convective downflows, delineates the boundaries of large scale cells of overturning plasma and exhibits “voids” in magnetic organization. 
Such voids include internetwork fields, a mixed-polarity sparse field that populate the inner part of network cells.\\ 
To single out voids and to quantify their intrinsic pattern a fast circle packing based algorithm is applied to 511 SOHO/MDI high resolution magnetograms acquired during the outstanding solar activity minimum between 23 and 24 cycles.\\
The computed Void Distribution Function shows a quasi-exponential decay behavior in the range 10-60~Mm. 
The lack of distinct flow scales in such a range corroborates the hypothesis of multi-scale motion flows at the solar surface. 
In addition to the quasi-exponential decay we have found that the voids reveal departure from a simple exponential decay around 35~Mm.} 


\maketitle

\section{Introduction} 
The nature of the multi-scale magnetic pattern observed on the solar surface remains a long-standing puzzle in solar physics. 
The instability responsible of the solar convective zone is due to the increase of opacity in the outer envelope of the star. 
The resulting turbulent convection creates temperature and velocity structures which evolve over a range of spatial and temporal multiple scales.
Large scale plasma motions (i.e. differential rotation, meridional circulation and torsional oscillations) and solar magnetic field contribute to increasing both the complexity of the solar surface dynamics and its appearence. 
Since the convective plasma flows govern the motion of the single magnetic features, the analysis of the photospheric magnetic pattern provides a way to investigate all the organization scales of convection, from the granulation to the global circulation \citep[e.g.][]{2011ApJ...727L..30Y}.

On the solar surface, three different spatial and temporal scales are traditionally identified: the granulation ($\sim 1$ Mm wide, few minutes lifetime), the meso-granulation ($5-10$ Mm, few hours lifetime), and the super-granulation ($30-50$ Mm, $\simeq 1$ day lifetime).
As reported by \citet{2009LRSP....6....2N} such a defined division is probably of historical origin, rather than physical. 
Indeed, all the analyses of the photospheric motion spectrum indicate that it covers all scales, from the global down to the sub-granular. Different image sampling and identification techniques, using different windows in space and time domains, may be responsible for the perceived different spatial patterns and different dominant scales.\\
In more detail, from the 60s on \citep{Leighton62,Leighton64}, the super-granulation has been the subject of a number of studies about 
its origin \citep{2003ApJ...597.1200R,2007ApJ...662..715C,2007AIPC..948..111S}, 
its spatial scale \citep{1997ApJ...475..328S,1998SoPh..180...29,2004SoPh..221...23D,2008A&A...488.1109M}, 
its structure and evolution \citep{2004ApJ...616..1242, 2007A&A...472..599D,2009ApJ...707...67G,2010ApJ...725L..47D}, 
its large-scale organization \citep{2004SoPh..221...33B,2005ApJ...632..677B,2004ApJ...608.1167L}, and 
its relationship with the surface magnetic fields \citep{2000SoPh..197...21L,2000A&A...357.1063R,2012ApJ...758L..38O}, 
just to cite a few. 
A detailed review of these efforts is presented in \citet{2010LRSP....7....2R}.\\
Although some differences between the supergranulation pattern and the magnetic network have been noticed, \citep[e.g.:][]{2010LRSP....7....2R}, the general conclusion is that convective cells at super-granular scale are outlined by the magnetic network, the large-scale interconnected pattern of magnetic elements.
With the increase of the spatial resolution of instrumentation over the last few years, it has been made clear that the so called ``quiet Sun'' is in fact crowded with magnetic flux structures. 
Their magnetic field strengths ranging from below 200 to 1000 Gauss.

The inspection of photospheric magnetograms, at the limits of the available resolution, \citep{2003A&A...411..615S,2008ApJ...672.1237L,2012ApJ...755.175} reveals the presence of regions "empty" of magnetic elements, commonly named \textit{voids}, whose spatial distribution and pattern is due to underlying plasma flow structures. 

In this paper we study the \textit{void} size distribution, to check whether it reveals distinct flow scales (e.g. supergranular scale) or it could be described by a smooth distribution.
\\
The automatic and accurate identification of voids, necessary to quantify their geometrical properties, is performed using an improved version of the void detection algorithm introduced in \citet{2013SoPh..282..379B} (henceforth Paper~1). A novel and faster circle packing algorithm has been developed and applied to identify such voids in the 10-60~Mm spatial range of SOHO/MDI high resolution 'quiet sun' magnetograms acquired during the last solar activity outstanding minimum. 
We stress that the analysis of void statistics, as derived from the Void Size Distribution (VSD), could be a powerful tool to describe the magnetic pattern on the photosphere and to possibly test current and future solar MHD numerical simulations.

\section{Dataset}
\label{S-dataset} 
The analysis is performed on an extended series of SOHO/MDI magnetograms acquired during the solar activity minimum between cycles 23 and 24. The Solar Oscillations Investigation/Michelson Doppler Imager (SOI/MDI) instrument \citep{1995SoPh..162..129S}, on board the SOHO spacecraft, acquired images of the Sun from 1996 to 2011 using the spectral line Ni I 676.78 nm \citep{2001SOI_TN_01_144,2011ApJ...728...92C}.
The images were recorded by a 1024$\times$1024 CCD camera, in two spatial resolution modes: full disk and high-resolution of the central part of the disk (HR).\\
We have analyzed a dataset of 511 high resolution magnetograms selected to cover a period of 18 months from Jan 1st, 2008 to Jun 30th, 2009. 
We carried out our analysis on the Level 1.8 high resolution magnetograms and we excluded from our dataset a few images which show strong magnetic activity. 
The images have a field of view of 11'$\times$11' with a plate scale of 0.625" per pixel and a (diffraction-limited) resolution of 1.25".
(see http://soi.stanford.edu/sssc/progs/mdi/calib.html for more information).\\

Such magnetograms measure the total amount of the longitudinal component of the magnetic flux density in each single resolution element: $\left \langle\mid \vec{B}\mid cos\gamma\right\rangle$, averaged over the pixel, where $\gamma$ is the angle between the magnetic vector $\vec{B}$ and the line of sight. In this work, we assume the LOS component of the magnetic field to be normal to the solar surface. We hereafter refer to $\mid \vec{B}\mid < cos\gamma >$ as B.\\

\begin{figure}[h!]
   \centering
   \includegraphics[scale=0.35,angle=0]{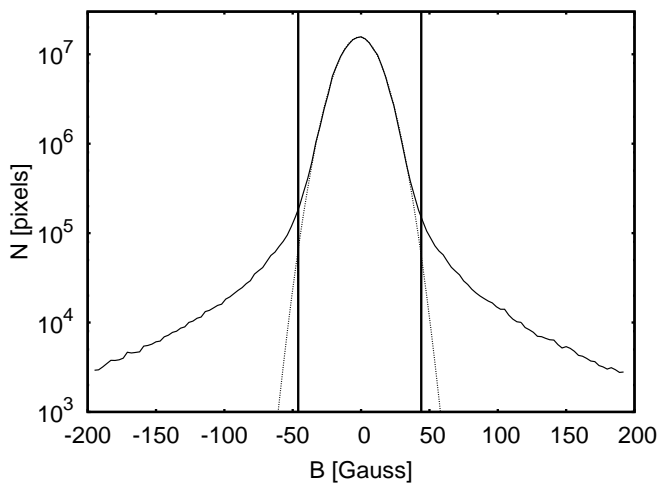}
   \caption{Histogram of the magnetic flux density signal from the whole dataset. The Gaussian fit highlights the noise character of the core of the distribution. The vertical continuus lines mark the $3\sigma$ interval}
   \label{noise}%
\end{figure}

Figure \ref{noise} shows the histogram of the magnetic signal covering the whole dataset: its core is well approximated by a Gaussian distribution 
 \citep[as in e.g.][]{2004A&A...417.1125K}. 
The standard deviation of the magnetic signal measured from our dataset is $\sigma \simeq$ 15 Gauss and the vertical lines mark the $3\sigma$ interval around the mean of signals 

\section{Void detection algorithm}
\label{S-algorithm2} 

To analyze the large number of selected magnetograms we needed an improved and faster version of the void-searching method already used in Paper~1.
A detailed description of such method has been given by \citet{1998ApJ...497..534A}, henceforth AM98, and in the same Paper~1. 
We achieved a remarkable improvement of performances by modifying the most time consuming part of the AM98 method: the \textit{climbing algorithm}. 
In AM98, each non-magnetic pixel in the binarized image has to be addressed to its proper void through the construction of a continuously connected path ``along a monotonically increasing $\nabla D(x)$'' (i.e. the climbing algorithm), $D$ being the \textit{distance field} generated from the binarized magnetogram. 

\begin{figure}[h!]
   \centering
	   \includegraphics[scale=0.3,angle=0]{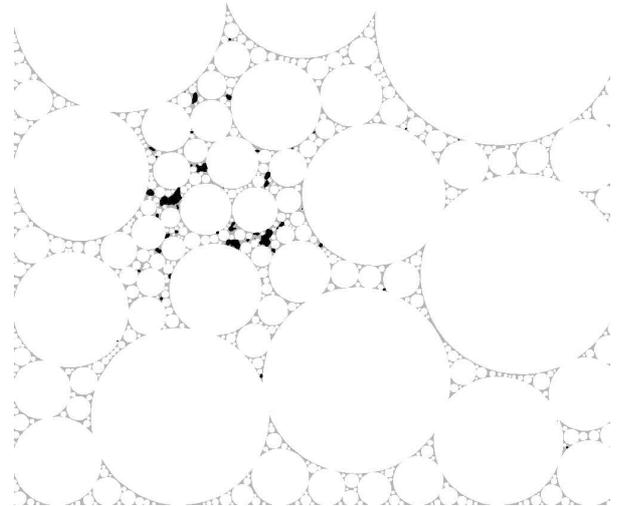}
   \caption{Circle packing in a binarized magnetogram: the magnetic structures (dark) are surrounded by circles which cover almost completely the non-magnetic area (grey).}
   \label{bubbles_ex}
\end{figure}

In the new version of the void detection algorithm we apply a recursive circle packing procedure filling the non-magnetic area of the binarized magnetogram. 
The climbing algorithm is then executed only for the centres of the filling circles, saving a noteworthy amount of CPU time. 
The recursive packing (figure~\ref{bubbles_ex}) is done through the generation of a sequence of modified distance fields (DFs) relative to the original binarized image. In each step a circle is drawn centered in the maximum of the current DF and having radius equal to the local DF value (i.e. extending the radius up to the \textit{touch} with the nearest magnetic structure). In the following step we insert the last generated circle as a fictitious magnetic area in the previously used binarized image and we calculate a new DF.

Likewise in the Apollonian circle packing \citep{2011ExpMath...20.380}, a sequence of circles ${C_1,C_2,C_3,...}$ having non-crescent diameters and filling the regions ''empty'' of magnetic elements is recursively generated. The iteration is stopped when the current $C_i$ radius $R$ is smaller than an assigned $R_{min}$. 

With the new algorithm we reduce the time required for the void identification (about 25 times shorter than in AM98).

\begin{figure*}[htb]
   \centering
   \includegraphics[scale=0.32]{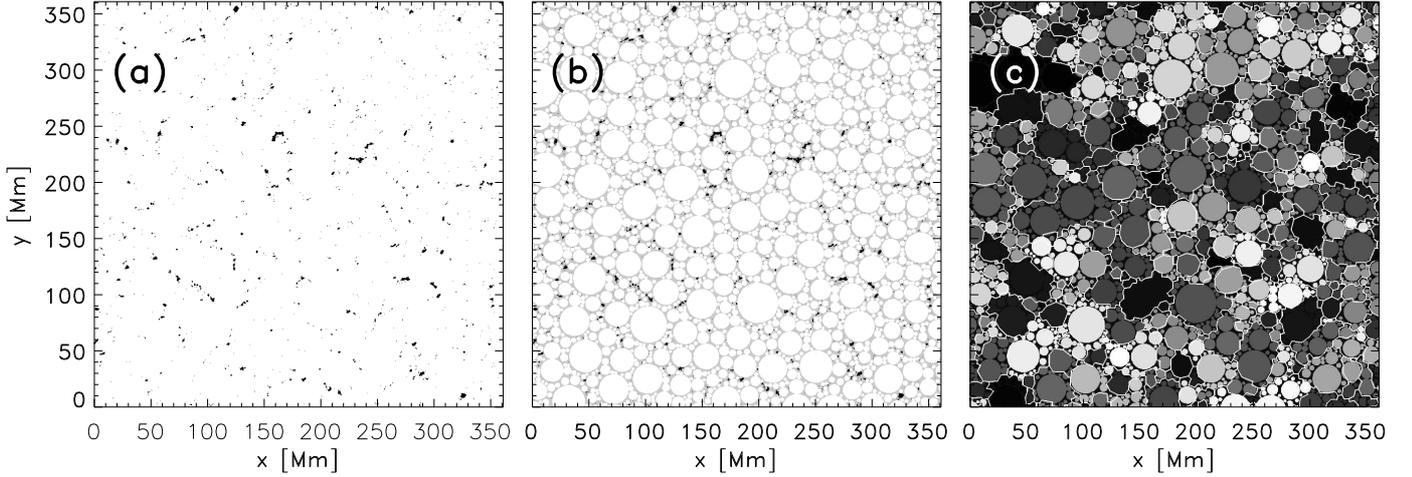}
   \caption{
   		Panel (a), the detected magnetic structures (in black) on a typical binarized SOHO/MDI magnetograms are reported. The threshold is equal to $3\sigma$; Panel~(b) the circles obtained by recursively defining the local DF maxima;Panel (c) the circles are packed together to generate the voids. All the circles pertaining to the same void structure are labeled with the same index, i.e. color (in this figure we highlight voids with a white boundary). The panels show a portion of the MDI FOV for the sake of visualization.}
              \label{algorithm}%
    \end{figure*}
     
In figure~\ref{algorithm} an example of the circle packing coverage relative to one of the magnetogram of our dataset is shown.
Panel~(a) shows a the binarized magnetogram. 
On the whole dataset we applied a segmentation fixed threshold of $45$ Gauss, which roughly corresponds to $3\sigma$ (see figure~\ref{noise}). 
Panel~(b) displays the circle packing coverage relative to the same magnetogram.
Finally, the Panel~(c) of the same figure~\ref{algorithm} shows the result of this circle addressing process:
all the circles pertaining to the same void structure are labeled with the same index, i.e. color. 
In this figure the groups of circle belonging to the same void are enclosed by a white boundary only to highlight them.

It is important to point out that the total area of circles used to identify the voids do not fill entirely the area of a magnetogram. This is due to the cutoff imposed by the smallest circles (in this work we set $R_{min}=2~pixel \simeq 1$ Mm). 
A careful analysis of a sub-set of magnetograms, analyzed using both the classical AM98 algorithm and the circle packing procedure, allowed us to compute the factor (0.79) which corrects the effect due to circles partial coverage.


\section{Analysis and results}
\label{S-results} 

\begin{figure}[h!]
   \centering
	   \includegraphics[scale=0.9,angle=0]{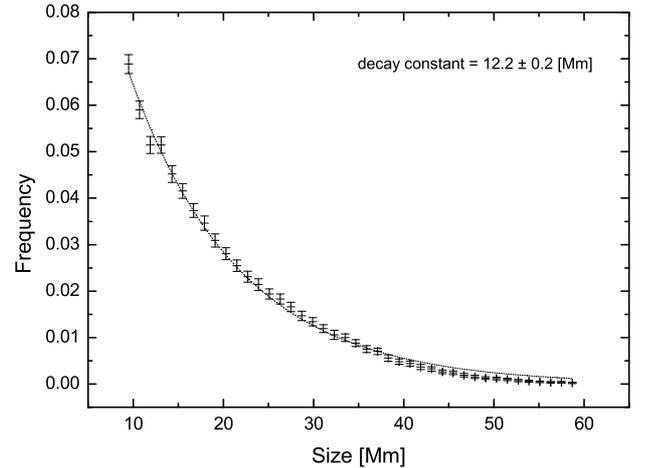}
   \caption{Void Size Distribution of 252488 voids singled out in the whole SOHO/MDI dataset. Error bars are computed via a 1000-samples bootstrap on 511 magnetograms. The dotted line represents the exponential decay fit, whose parameters are shown in the upper right corner.}
   \label{linexpdecay}%
\end{figure}

The void detection algorithm singled out 252488 voids from the 511 SOHO/MDI magnetograms. 
In order to derive the distribution of the void sizes (VSD), we calculated the area $A$ in Mm$^2$ of the observed voids and defined their \textit{Size} as the diameter of a circular region with the same area, Size=$2 \sqrt{A/\pi}$.
\\
The VSD is shown in figure \ref{linexpdecay} together with an exponential fit $ F=A \times e^{-Size/S_{d}}$ where $S_{d}$ is the decay constant in Mm. 
\\
The error bars are calculated as the standard deviation of 1000-resampling ensembles obtained randomly extracting sub-samples of 60 magnetograms from the complete dataset (bootstrapping method). 
They highlight also the remarkable stability of the VSD during the $\sim$1.5~year of observations, which, we recall, have been acquired during the period of minimum solar activity between cycles 23 and 24.\\
Slightly modifing the threshold at 45 Gasuss does not affect the shape of the void size distribution (see Paper1).\\
There are two important outcomes from the VSD shown in figure \ref{linexpdecay}. 
First, the computed VSD depends strongly on the size of voids. 
Indeed, the distribution is in a very good agreement (R$^2$=0.996) with an exponential decay with constant $S_{d}=12.2\pm0.2$ Mm in the range 10-60 Mm. 
Second, no particular feature is observed around the supergranular $30-50$ Mm scale. 
In other words, the VSD appears to rule out supergranulation as a distinct scale of motion, supporting the multi-scale nature of convective ``motion flows'' at the solar surface \citep{2009LRSP....6....2N, 2011ApJ...727L..30Y, 2013SoPh..282..379B}.\\

\begin{figure}[h!]
   \centering
   \includegraphics[scale=0.35,angle=0]{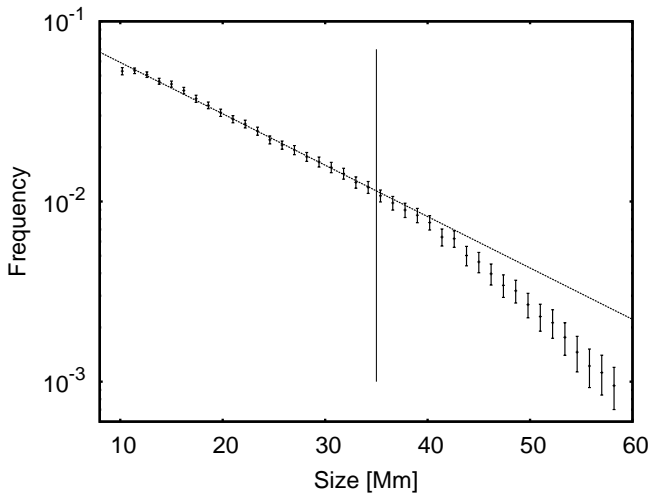}
   \caption{Symbols with error bars represent the Void Size Distribution of 252488 voids in semilog scale. Exponential fits in the regions 10-35 Mm and 35-60 Mm are represented by dotted lines. In the range 35-60 Mm the decay constant is $S_{d}=7.6\pm0.2$ Mm while in the range 10-35 Mm the value is $S_{d}=12.4\pm0.2$ Mm. The continuous vertical line marks the 35Mm scale.}
   \label{bootstrap}%
\end{figure}

\begin{figure}[h!]
   \centering
   \includegraphics[scale=0.35,angle=0]{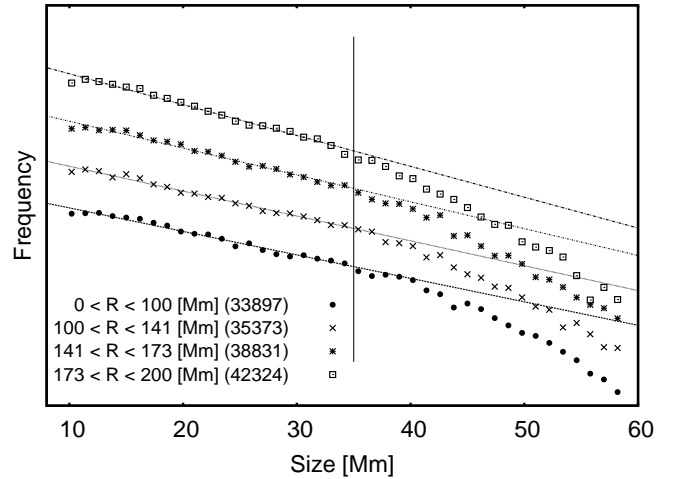}
   \caption{VSDs for subsets of voids extracted in four annuli centered on the solar disc center. The distributions corresponding to the four regions are represented by the different lines. Each of the subsets keeps the regime modification at around 35~Mm indicated by the vertical line. The number between brackets indicates the number of voids found in each annulum.
Baseline offsets are adjusted for each VSD in order to separate the distributions.}
   \label{anelli}%
    \end{figure}

The standard exponential decay satisfactorily fits the distribution, but it also seems to overestimate the size of the largest voids. 
Indeed, a careful analysis shows that there is a regime change around 35~Mm. 
Actually, the VSD reveals, in a semilog plot (figure \ref{bootstrap}), a departure from a simple exponential decay around 35~Mm. 
In order to determine if the departure from the exponential decay is due to projection effects or inhomogeneities on the SOHO/MDI field of view of high-resolution images (about $1/3$ of Sun's diameter) we checked the possibility of a systematic center-to-limb effect in the detection of voids. 
Figure \ref{anelli} shows the VSD in four annular regions centered on the FOV center. 
The radii are chosen to keep constant each region area and each void is allocated to one of the annular region according to the position of its baricenter. 
In the figure baseline offsets are adjusted for each VSD in order to separate the distributions.
We observe a slight trend in the void distributions and the growth of small size voids on moving away from the magnetogram center. 
Despite this global trend, each of the sub-samples keeps the departure from a simple exponential decay around 35~Mm. 
As a consequence, it would seem reasonable to draw a conclusion that projecton has a detectable effect on VSD, probably due to the different optical depth of the designed regions, but the regime modification around 35 Mm is a real feature of solar (turbulent) convection.

\begin{figure}[h!]
   \centering
   \includegraphics[scale=0.45,angle=0]{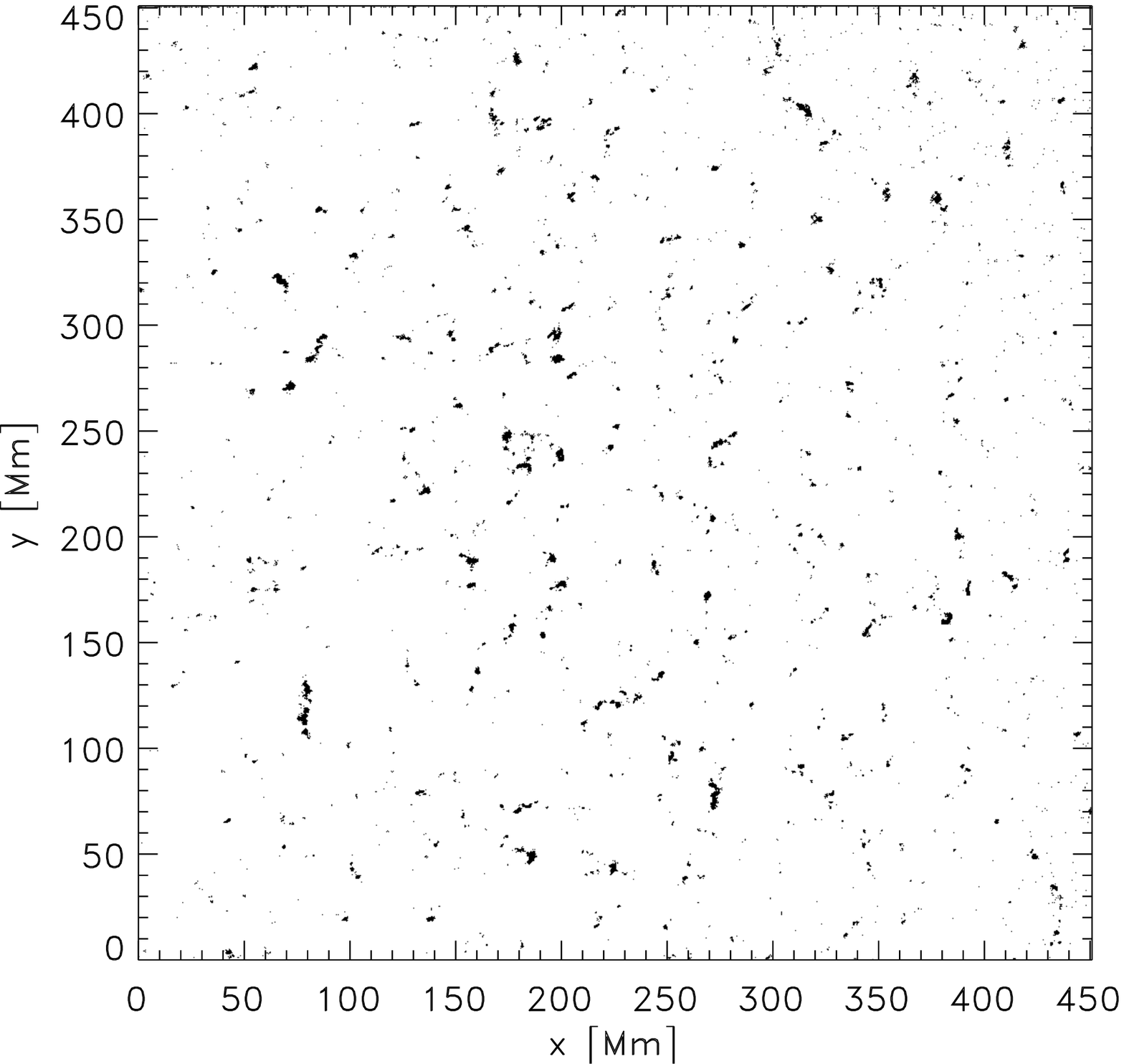}
   \includegraphics[scale=0.45,angle=0]{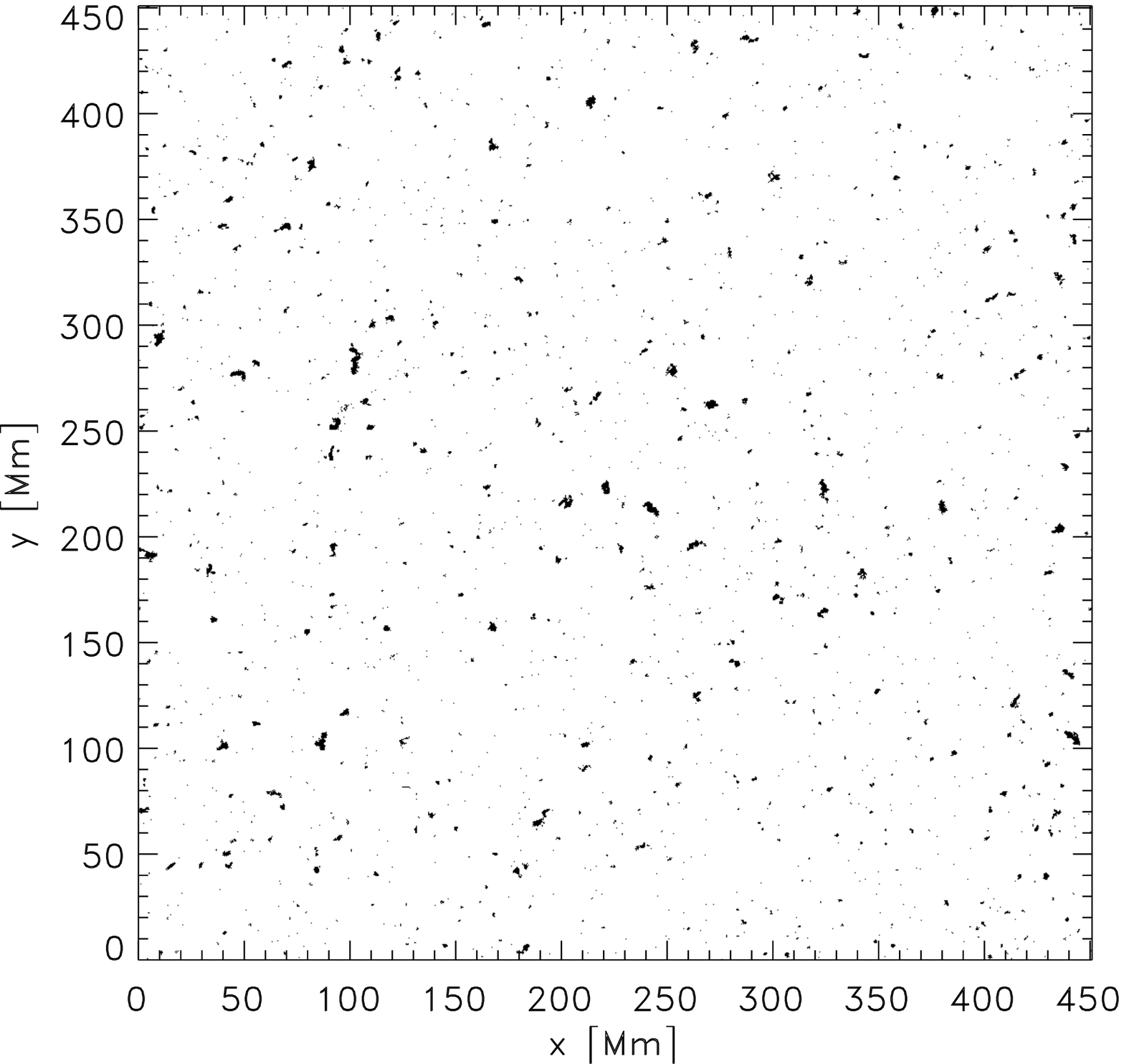}
   \caption{The upper panel gives the MDI magnetogram (July 11, 2008) while the lower panel results after shuffling the original magnetic structures.}
   \label{shuffle2}%
\end{figure}

\begin{figure}[h!]
   \centering
   \includegraphics[scale=0.35,angle=0]{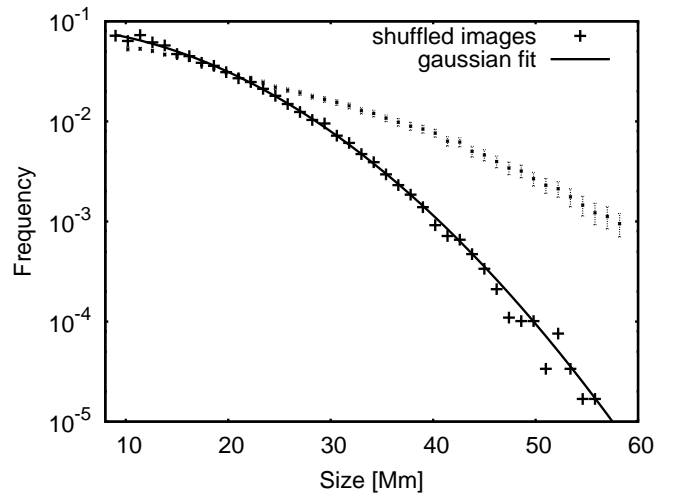}
   \caption{Symbols with error bars represent the Void Size Distribution of 252488 voids (as in Fig. \ref{linexpdecay}) in semilog scale. Crosses show the mean VSD computed shuffling the magnetic structures in each magnetogram. The Gaussian fit is represented by the continuous curve.}
   \label{gaussian}%
\end{figure}
A basic further question about the sparse pattern of magnetic structures, as singled out by the thresholding in our method (figure \ref{algorithm}[a]), is whether such structures are distributed in a random fashion or they are somehow ordered.
The same question can be expressed in terms of VSD: is it possible to obtain a similar distribution from a random distributed pattern of magnetic structures?
\\  
We investigated this possibility by applying the void detection algorithm on {\it shuffled} images. 
Such images were obtained by shuffling in the field of view the magnetic structures of each binarized magnetogram. Such a reshaping  was performed by avoiding the superimposition of the translated structures, i.e. preserving the total number of magnetic pixels in each binarized image. The original and shuffled images corresponding to MDI magnetogram acquired on July 11, 2008 are shown for a comparison in figure \ref{shuffle2}.\\

The resulting VSDs is shown in the semilog plot of figure \ref{gaussian} (crosses), together with the original VSD (symbols with errorbars). In this case, the shuffling was applied on the images of the second half of 2008 (167 magnetograms). 
In the same figure, a Gaussian fit (continuous line) of the VSDs is shown. 
The evident difference between the two distributions strongly suggests that the VSD obtained from SOHO/MDI magnetograms is not compatible with a simple random-like pattern as discussed in
\citet{2003A&A...402..1115,2004SoPh..221...33B,2008SoPh..251...417}.

\section{Conclusions}
\label{S-conclusions} 
In this paper, we analyzed the distribution of void sizes in quiet Sun magnetograms to investigate the multi-scale magnetic pattern on the photosphere, therefore the organization scales of convection \citep{2011ApJ...727L..30Y}, in the 10-60~Mm range.\\
In particular, we have investigated the geometrical properties of voids of magnetic field, associated to the boundaries of large scale cells of overturning plasma using 511 SOHO/MDI high resolution magnetograms acquired during the last outstanding solar activity minimum between 23 and 24 cycles.
We selected this period to minimize the influence of active regions on the spacial pattern due to convective motions.
Our study is based on a fast void detection procedure which uses a circle packing algorithm to obtain robust statistics.\\
\\
During the 1.5 year covered by the dataset the computed VSD shows a remarkable stability.
There is a slight dependence of the VSD from the curvature of the field of view, i.e. from the position of voids in respect of the angle between the line of sight and the solar surface vertical. It is evident that such an angle interferes with the magnetograms segmentation with a fixed  and constant threshold. 
On considering the magnitude of the effect that we have observed in our dataset, it was beyond the scope of this work to deal with any segmentation improvement, for the moment.
The observed stability has the potential to be used as a quantitative definition of ''quiet Sun'' as well as of ''solar activity minimum''.\\        
\\
We have established that the VSD shows a quasi-exponential decay in the observed range. The monotonic distribution and the lack of marked features in the 10-60~Mm range point out the absence of a supergranular scale and support the multi-scale hypothesis of convective motion flows at the solar surface \citep{2009LRSP....6....2N, 2011ApJ...727L..30Y, 2013SoPh..282..379B}: however it does not rule out the reality of convection structures with those sizes.
For example, an advective-interaction model \citep{2003ApJ...597.1200R} is able to explain the presence of flow at large scales as the result of local merger of downward plumes.\\
\\
In addition to the quasi-exponential decay we have found that the voids reveal departure from a simple exponential decay around 35~Mm.\\
We speculate about a possible scenario leading to the existence of a convective instability scale. Namely, instabilities in the convection boundary layer located in the photosphere could be induced when the distance between downward plumes exceeds the above-mentioned critical value of 35~Mm. 
At this super-granular scale buoyancy forces becomes sufficient to trigger the formation of a new downflow plume with an instability mechanism similar to that reported in \citet{1995ApJ...443.863} for granulation scale. 
This mechanism would be able to introduce thermal plumes in the stratified compressible plasma providing possible sites for magnetic field concentrations. 
These new magnetic regions could set a limit to the maximum size of voids thus reducing the number of observed voids with typical sizes greater than 35~Mm.

%
\begin{acknowledgements}
We thank Stuart Jefferies and Juri Toomre for the stimulating discussions.
SoHO is a mission of international cooperation between ESA and NASA. This work was supported in part by the EC Home Affairs CIPS program SPARC: \textit{Space Awareness for Critical Infrastructure} research grant. This research work is partly supported by the Italian MIUR-PRIN grant 2012P2HRCR on "The active Sun and its effects on Space and Earth climate" and by Space Weather Italian COmmunity (SWICO) Research Program.
\end{acknowledgements}

\mbox{}~\\

\end{document}